\documentstyle[sprocl]{article}

\input{psfig}
\def\be{\begin{equation}}
\def\ee{\end{equation}}
\def\bea{\begin{eqnarray}}
\def\eea{\end{eqnarray}}

\def\lesssim{\hbox{$ {     \lower.40ex\hbox{$<$}
                   \atop \raise.20ex\hbox{$\sim$}
                   }     $}  }
\def\gtrsim{\hbox{$ {     \lower.40ex\hbox{$>$}
                   \atop \raise.20ex\hbox{$\sim$}
                   }     $}  }

\def\mycite{\,\cite}

\begin{document}

\title{GRAVITATIONAL WAVES FROM COSMIC STRINGS}

\author{R.A. Battye}

\address{Theoretical Physics Group, Blackett Laboratory, Imperial College\\
Prince Consort Road, London SW7 2BZ, U.K.}

\author{R.R. Caldwell}

\address{Department of Physics and Astronomy,  
University of Pennsylvania \\ 209 South 33rd Street, Philadelphia, PA 19104-6396,
U.S.A.}

\author{E.P.S. Shellard}

\address{Department of Applied Mathematics and Theoretical Physics, University of
Cambridge \\ Silver Street, Cambridge CB3 9EW, U.K}


\maketitle\abstracts{Gravitational waves from cosmic strings are generated in the first fractions of a second after the Big Bang, potentially providing a unprecedented probe of the early universe. We discuss the key dynamical processes underlying calculations of the stochastic background produced by a string network and we detail the parameter dependencies of the resulting spectral density $\Omega_{\rm gr}(f)$. The present constraints on the cosmic string mass scale $\mu$ arising from the millisecond pulsar timings and primordial nucleosynthesis are discussed, justifying our present conservative bound 
$G\mu/c^2 < 5.4(\pm 1.1)\times 10^{-6}$. We then discuss the strong prospect of detecting (or ruling out) cosmic string background with the next generation of gravitational wave experiments. Comparison is also made with alternative cosmological sources of gravitational waves such as inflation and hybrid topological defects.}

\section{Introduction}

An outstanding achievement of modern cosmology has been the detection
of anisotropies in the cosmic microwave background\mycite{smoot}.
These anisotropies provide a snapshot of the universe about 400,000
years after the Hot Big Bang,  just as the universe became transparent
to electromagnetic radiation. Despite  this observational triumph, as
yet it has  been insufficient to differentiate between competing
paradigms for galaxy  formation, nor has it shed much light on
cosmological processes taking  place before photon decoupling. There
are, however, other types of radiation. Gravitational radiation, as
considered in the present work, may penetrate through this
electromagnetic surface of last scattering, and travel virtually
unaffected since emission. Of course, for gravitons this remarkable
transparency is due to their very weak interaction  with ordinary
matter which, in turn, makes them difficult to observe. However,
pioneering experiments have been proposed which  could detect a
stochastic background of gravitational waves generated  in the early
universe over a range of frequencies.

Cosmic strings are line-like topological defects which may have formed
during a phase transition in the early
universe\mycite{Historicref,Reviewref}.  Strings which formed with a
mass-per-unit-length $\mu$ such that $G\mu/c^2 \sim 10^{-6}$ may be
responsible for the formation of the large-scale structure and cosmic
microwave background anisotropy observed in the universe today. In
general, a network of strings will evolve towards a self-similar
scaling regime by the production of loops and subsequent emission of
gravitational radiation. Since the sources of these gravitational waves
come from many horizon volumes this background will be stochastic and
is likely to appear as noise on a gravitational wave detector.

A stochastic background of gravitational waves is normally quantified
by the relative spectral density $\Omega_{\rm g}(f)$ given at a
frequency $f$. That is, the energy density in gravitational radiation
in  an octave frequency bin centred on $f$, relative to the critical
density of the universe. This is directly related to the dimensionless
wave amplitude  ($h_c \propto \sqrt{\Omega_g}/f$) which is measured
experimentally.

In the case of gravitational radiation produced by a cosmic string
network, for a given frequency today we may identify  a characteristic
time at which the waves were emitted. Assuming that the radiation is
emitted at a time $t_{\rm e}$ before equal matter-radiation ($t_{\rm
e}<t_{\rm eq}\sim 4,000$ years) and that it is created with a
wavelength comparable to the horizon $\lambda(t_{\rm e})\sim t_{\rm
e}$, then the frequency today is given  by $f \sim z_{\rm
eq}^{-1}(t_{\rm eq}t_{\rm e})^{-1/2}$ where the redshift  is $z_{\rm
eq} \sim 2.3\times10^{4}\Omega_0 h^2$. In particular those
gravitational waves created at the electroweak phase transition will
have a frequency $f\sim 10^{-3}{\rm Hz}$, whereas those created at the
time of radiation-matter equality will have a frequency $f\sim
10^{-13}{\rm Hz}$.

The relevant constraints on stochastic backgrounds are due to the
measured timing residuals in pulsar signals and also from Big Bang
nucleosynthesis.  The most recent analysis of pulsar signal arrival
times\cite{MZVLref} gives the limit on the spectral density of gravitational
radiation
\begin{equation}
\Omega_{\rm gr}(f) \equiv
{f \over \rho_{\rm crit}}{d \rho_{\rm gr} \over d f} \mid_{f_{\rm obs}}\,\,
 <  9.3 \times 10^{-8} h^{-2}  \quad (95 \% \,{\rm CL})
\label{GRBLimits}
\end{equation}
in a logarithmic frequency interval at $f_{\rm obs} = (8 \, {\rm
yrs})^{-1}$.  This analysis corrects errors in previous
work\mycite{KTRref,TDref},  and uses an improved method for  testing
the hypothesis that the timing noise is due to a stochastic
gravitational wave background. For these reasons we use the latest
bound, equation (\ref{GRBLimits}), to obtain a constraint on $G\mu/c^2$
for given values of the cosmic string and cosmological parameters.

The bound from nucleosynthesis is slightly different, as it constrains
the entire spectrum. The standard constraint quoted is that the total
energy density in gravitational waves at the time of nucleosynthesis
must be less than 5.4\% of the critical density, or
\begin{equation}
\Omega_{\rm gr}=\int{d\omega\over\omega}\Omega_{\rm gr}(\omega)\approx
0.162(N_{\nu}-3)\Omega_{\rm r}<0.054\,,
\end{equation} 
which can be thought of as a contribution from an extra neutrino
species. This constraint is somewhat weakened at present, due to
uncertainties in the observed element abundances. For this reason, we
shall discuss constraints from a range of values for $N_{\nu}$, the
effective number of neutrino species.

In this proceedings, based in part on work in ref.[7], we review the status of the stochastic
background produced by cosmic strings. The details of the spectrum of
gravitational radiation due to cosmic strings has been previously
considered elsewhere\mycite{RadiationCalcref}.  Here we benefit from
recent work\mycite{BackReactionref}, which
suggests that the effect of the radiative back-reaction on a string
loop is to damp out the higher oscillation modes. We illustrate the
effects of a non-standard thermal history, low-$\Omega$ universes and
discuss hybrid systems of defects, such as strings and domain walls or
strings connected by monopoles.  Finally, we discuss the opportunities
for detection.

\section{Cosmic string loop radiation spectrum}
\label{EmissionModelSection}

The spectrum of gravitational radiation emitted by a network of cosmic
string loops is obtained using a background $\Omega = 1$ FRW
cosmological model, an extended one-scale model for the evolution of a
network of cosmic strings, and a model of the emission of gravitational
radiation by cosmic string loops. The procedure by which the spectrum
is computed has been presented in detail by Caldwell and
Allen\mycite{RadiationCalcref}. In this section we discuss the model
for radiation by an individual loop.

The model of the emission of gravitational radiation by cosmic string
loops is composed of the following three elements.
\begin{enumerate}
\item
A loop radiates with power $P = \Gamma G \mu^2 c$.  The dimensionless
radiation efficiency, $\Gamma$, depends only on the loop configuration,
rather than overall size. Recent studies of realistic
loops\mycite{CasperCommref} indicate that the distribution of values of
the efficiency has a mean value $\langle\Gamma\rangle \approx 60$.
\item
The frequency of radiation emitted by a loop of invariant length $L$ is
$f_n = 2 n/L$ where $n=1,2,3,...$ labels the oscillation mode.  
\item
The fraction of the total power emitted in each mode of oscillation $n$
at frequency $f_n$ is given by the coefficient $P_n$ where
\begin{equation}
P =  \Big(\sum_{n=1}^\infty P_n \Big) G \mu^2 c = \Gamma G \mu^2 c.
\label{EmissionModel}
\end{equation}
Analytic and numerical studies suggest that the radiation efficiency
coefficients behave as $P_n \propto n^{-q}$ where $q$ is the spectral
index.
\end{enumerate}
This model has several shortcomings. First, the spectral index $q$ has
not been well determined by the numerical simulations. Numerical
work\mycite{CuspyIndexref} suggests $q = 4/3$  as occurs with cuspy
loops, loops along which points momentarily reach the velocity of
light, based on simulations of a network of cosmic strings. However,
these simulations have limited resolution of the important small scale
features of the long strings and loops.  Hence, the evidence for
$q=4/3$ is not compelling. Analytic work\mycite{KinkyIndexref} suggests
that $q = 2$, characteristic of kinky loops, loops along which the
tangent vector changes discontinuously as a result of intercommutation,
may be more realistic. Second, the effect of back-reaction on the
motion of the string has been ignored. In this model, a loop radiates
at all times with a fixed efficiency, in all modes, until the loop
vanishes.  As we shall next argue, the back-reaction will result in an
effective high frequency cut-off in the oscillation mode number.  Thus,
the loop will only radiate in a finite number of modes, and hence in a
finite range of frequencies. The resolution of these issues may have
strong consequences for the entire spectrum produced by a network of
strings.

Recent advances in the understanding of radiation back-reaction on
global strings suggest various modifications to the simplified model of
emission by cosmic string loops.
It has been shown\mycite{DetailedRadiationCalcref}  that there are
remarkable similarities between gravitational radiation and Goldstone
boson radiation from strings, which we believe allow us to make strong
inferences as to the nature of gravitational radiation back-reaction.
For example, the same, simple model for the emission of gravitational
radiation by cosmic strings may be transferred over to global strings:
in the absence of Goldstone back-reaction,  global string loops radiate
at a constant rate, at wavelengths given by even sub-multiples of the
loop length, with an efficiency as described by an equation similar to
(\ref{EmissionModel}).  Hence, our argument proceeds as follows. Fully
relativistic field theory simulations of global strings have been
carried out\mycite{BackReactionref}, where it was observed that the
power in high oscillation modes is damped by the Goldstone
back-reaction on periodic global strings.  An analytic model of
Goldstone back-reaction\mycite{BackReactionref}, as a modification of
the classical Nambu-Goto equations of motion for string, was developed
which successfully reproduces the behaviour observed in field theory
simulations. That is, high frequency modes are damped rapidly, whereas
low frequency modes are not.  Thus, we are motivated to rewrite
equation (\ref{EmissionModel}) for global strings, and by analogy for
cosmic strings, as 
\begin{equation} P=\Big( \sum_{n=1}^{n_*}P_n \Big) G
\mu^2 c =\Gamma G\mu^2 
\end{equation} 
where $n_*$ is a cut-off\mycite{VV} introduced to incorporate the effects of
back-reaction. By comparing the back-reaction length-scale to the loop
size, we estimate that such a cut-off should be no larger than $\sim
(\Gamma G \mu/c^2)^{-1}$.  The ongoing investigations of global and
cosmic string back-reaction\mycite{BackReactionResearchref} have not
yet reached the level of precision where a firm value of $n_*$ may be
given. As we demonstrate later, the effect on the radiation spectrum is
significant only for certain values of the cut-off.

\section{Analytic estimate of the radiation spectrum}
\label{AnalyticEstimateSection}

Analytic expressions for the spectrum of gravitational radiation emitted by a
network of cosmic strings have been derived elsewhere\mycite{RadiationCalcref,Reviewref}. While
these analytic expressions are simplified for convenience, they offer the
opportunity to examine the various dependencies of the spectrum on cosmic string and
cosmological parameters.

The spectrum of gravitational radiation produced by a network of cosmic strings has
two main features. First is the `red noise' portion of the spectrum with nearly
equal gravitational radiation energy density per logarithmic frequency interval,
spanning the frequency range $10^{-8}\, {\rm Hz} \lesssim f \lesssim 10^{10}\,{\rm
Hz}$. This spectrum corresponds to gravitational waves emitted during the
radiation-dominated expansion era. This feature of the spectrum may be accessible to
the forthcoming generation of gravitational wave detectors. Second is the peak in
the spectrum near $f \sim 10^{-12} \, {\rm Hz}$. The amplitude and slope of the
spectrum from the peak down to the flat portion of the spectrum is tightly
constrained by the observed limits on pulsar timing noise.

\subsection{Red noise portion of the spectrum}

An analytic expression for the `red noise' portion of the gravitational
wave spectrum is given as follows:
\begin{equation}
{f \over \rho_{\rm crit}}{d \rho_{\rm gr} \over d f}
= {8 \pi \over 9} A {\Gamma(G\mu)^2 \over \alpha c^4}
\bigl[1 - {\langle v^2\rangle / c^2}\bigr] {(\beta^{-3/2} - 1) \over (z_{\rm
eq} + 1)}
\qquad 10^{-8}\, {\rm Hz} \lesssim f \lesssim 10^{10}\,{\rm Hz}
\label{RedNoiseSpectrum}
\end{equation}
\begin{equation}
A  \equiv   \rho_{\infty} d_H^2(t) c^2 / \mu \qquad\qquad
\beta  \equiv   \bigl[1 + f_{\rm r}\,\alpha d_H(t) c/(\Gamma G \mu
t)\bigr]^{-1}.
\end{equation}
In the above expressions, $\rho_{\infty}$ is the energy density in `infinite' or
long cosmic strings, $\alpha$ is the invariant length of a loop as a fraction of the
physical horizon length $d_H(t)$ at the time of formation, $\langle v^2 \rangle$ is
the rms velocity of the long strings, and $f_{\rm r}\approx 0.7$ is a correction for
the damping of the relativistic center-of-mass velocity of newly formed string
loops. All quantities are evaluated in the radiation era; $d_H(t) = 2 c t$, $A = 52
\pm 10$, and $\langle v^2 \rangle/c^2 = 0.43 \pm 0.02$ as obtained
from numerical simulations\mycite{Simulationref}.

\begin{figure}
\psfig{figure=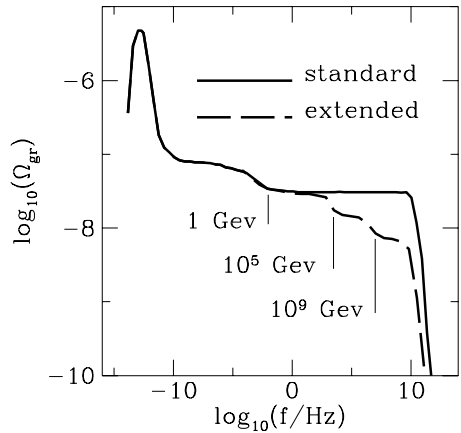,height=7.0cm} 
\caption{
The effect of a non-standard thermal history of the cosmological fluid on the
amplitude of the red noise portion of the gravitational wave spectrum is shown. The
solid curve displays the spectrum produced using a minimal GUT with a maximum $g =
106.75$.  The dashed curve shows the spectrum produced allowing for a hypothetical,
non-standard evolution of $g(T)$, as might occur if there were a series of phase
transitions, or a number of massive particle annihilations as the universe cooled.
For temperatures $T > 10^9 \,{\rm GeV}$, the number of degrees of freedom is $g =
10^4$.  For $10^5 \, {\rm GeV} < T < 10^9 \, {\rm GeV}$, $g = 10^3$.  
For $T<10^5 {\rm GeV}$, the standard thermal scenario is resumed.}
\label{figure1}
\end{figure}

The above expression for the spectrum has been obtained assuming no change in the
number of relativistic degrees of freedom, $g$, of the background
radiation-dominated fluid. However, the annihilation of massive particle species as
the cosmological fluid cools leads to a decrease in the number of degrees of
freedom, and a redshifting of all relativistic particles not thermally coupled to
the fluid. This has the effect of modifying the amplitude of the spectral 
density\mycite{BennettPapersref}, equation (\ref{RedNoiseSpectrum}), by a factor
$(g(T_f)/g(T_i))^{1/3}$ where $g(T_{i,f})$ is the number of degrees of freedom at
temperatures before and after the annihilations.  Using a minimal GUT particle
physics model as the basis of the standard thermal history, we see that the red
noise spectrum steps downwards with growing frequency.
\begin{eqnarray}
{f \over \rho_{\rm crit}}{d \rho_{\rm gr} \over d f}
&=& {8 \pi \over 9} A {\Gamma(G\mu)^2 \over \alpha c^4}
\bigl[1 - \langle v^2\rangle/c^2\bigr] {(\beta^{-3/2} - 1) \over (z_{\rm eq} +
1)}
\cr\cr
&&\times\cases{
1	&  \cr
\qquad
10^{-8}\, {\rm Hz} \lesssim f \lesssim 10^{-10}\alpha^{-1}
	\, {\rm Hz} & \cr\cr
(3.36/10.75)^{1/3} = 0.68 & \cr
\qquad
10^{-10}\alpha^{-1}\, {\rm Hz} \lesssim f \lesssim 10^{-4}\alpha^{-1}
	\, {\rm Hz} & \cr\cr
(3.36/106.75)^{1/3}= 0.32 & \cr
\qquad
10^{-4}\alpha^{-1}\, {\rm Hz} \lesssim f \lesssim 10^{8}
	\, {\rm Hz} & }
\label{RedNoiseSpectrumToday}
\end{eqnarray}
Hence, the red noise spectrum is sensitive to the thermal history of
the cosmological fluid. The locations of the steps in the spectrum are
determined by the number of relativistic degrees of freedom as a
function of temperature, $g(T)$. As an example, we present the effect
of a non-standard thermal history on the spectrum in Figure
\ref{figure1}. In this sample model, the number of degrees of freedom
$g(T)$ decreases by a factor of $10$ at the temperatures $T=10^9,\,
10^5,\, 1\, {\rm GeV}$. The effect on the spectrum is a series of steps
down in amplitude with increasing frequency; detection of such a shift
would provide unique insight into the particle physics content of the
early universe at temperatures much higher than may be achieved by
terrestrial particle accelerators. In the case of a cosmological model
with a thermal history such that $g(T_i) \gg g(T_f)$ for $T_i > T_f$,
all radiation emitted before the cosmological fluid cools to $T_i$ will
be redshifted away by the time the fluid reaches $T_f$. As we discuss
later, such a sensitivity of the spectrum to the thermal history
affects the nucleosynthesis bound on the total energy in  gravitational
radiation, and the opportunity to detect high frequency gravitational
waves.

\subsection{Peaked portion of the spectrum}

We now turn our attention to the peaked portion of the gravitational
wave spectrum. The shape of this portion depends  critically on the
model for the emission by a loop, presented in section
\ref{EmissionModelSection}. Specifically, if a loop emits
in very high modes, a significant portion of the loop energy maybe radiated
at high frequencies, well above the fundamental frequency of the loop.
Hence, the dominant behaviour of
the peaked portion of the spectrum is given by
\begin{eqnarray}
{f \over \rho_{\rm crit}}{d \rho_{\rm gr} \over d f}
 &\approx & \cases{
C_1 /f^{(q-1)}	&	$1 < q < 2$ \cr
C_2 /f		&	$q \ge 2$ } \cr\cr
&& {\rm for}\quad 10^{-12} \, {\rm Hz} \lesssim f \lesssim 10^{-8} \, {\rm Hz}.
\label{Peaked Spectrum}
\end{eqnarray}

\begin{figure}
\psfig{figure=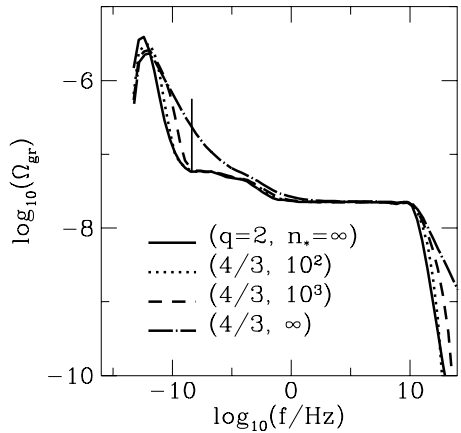,height=7.0cm}  
\caption{
The effect of a cut-off in the radiation mode number on the spectrum of
gravitational radiation is shown. Curves for the loop radiation
spectral index $q=2,\, 4/3$ for various values of $n_{*}$ are shown.
The vertical line shows the location of the frequency bin probed by
pulsar timing measurements. For $n_{*} {\protect{\lesssim}} 10^2$ the shape of
the
spectrum is insensitive to the value of $q$ for purposes of pulsar
timing measurements. For increasing $n_{*}$, more radiation due to
late-time cosmic string loops is emitted in the pulsar timing frequency
band. }
\label{figure2}
\end{figure}

\noindent Here $C_{1,2}$ are dimensionful quantities which depend on $G\mu/c^2,
\, \alpha, \,\Gamma, \, A, \, q$ and $n_{*}$. A lengthy expression
displaying the full dependence of the spectrum on these parameters is
not particularly enlightening, they can be found in ref.[13].The qualitative behaviour is as follows.  The overall
height of the spectrum depends linearly on $G\mu/c^2$, while the
frequency at which the peaked spectrum gives way to the red noise
spectrum depends inversely on $\alpha$.  The important result is that
for values of the mode cut-off $n_{*} \lesssim 10^{2}$, the spectrum
drops off as $1/f$ for any value $q \ge 4/3$. As a demonstration,
sample spectra with various values of $n_{*}$ are displayed in Figure
\ref{figure2}. Hence, the introduction of a sufficiently low mode
cut-off eliminates the dependence of the spectrum on $q$, the loop
spectral index.

\subsection{How stochastic is the background?}

We now comment on the statistical properties of the background. We would
like to know how good an approximation it is to claim that the signal 
is stochastic. The relevant quantity to consider is
the number of string loops which contribute to the radiation produced in a
particular frequency bin. If we identify, for each frequency $f$ observed, a unique
time of emission corresponding to the time $t_{\rm f}$ that the loop formed, then
one can calculate the number of individual horizon cells on the sky contributing to the
radiation background at each frequency:
\begin{equation}
N(f)\approx 4\times 10^{17}\alpha^2\left({f\over f_{\rm p}}\right)^2.
\end{equation}
Here, $f_{\rm p}\approx 4\times 10^{-9}{\rm Hz}$ is the frequency corresponding to
horizon sized gravitational waves that would be detected by the milli-second pulsar
and $\alpha$ is the loop production size with respect to the horizon.
For the frequencies of interest, $f\approx 100{\rm Hz}$ (LIGO/VIRGO), $f\approx
10^{-3}{\rm Hz}$ (LISA) and $f\approx 4\times 10^{-9}{\rm Hz}$ (pulsar),  the values
of $N$ are $10^{28}$, $10^{18}$ and $4\times 10^{7}$ respectively, where we have
used $\alpha\sim 10^{-5}$ for illustrative purposes. Hence, we may argue from the
central limit theorem, that if the $N$ cells act independently, the sum of the
amplitudes of radiation arriving at the detector or pulsar is gaussian and the
background is stochastic. This gaussian assumption may break down at a frequency of
$10^{-12}{\rm Hz}$ where $N\sim 1$, corresponding to radiation emitted around a
redshift $z\sim 4$. In this case, the radiation is due to a relatively small number of
sources which are on average a few hundred megaparsecs away. These sources would
appear as burst sources at a detector, rather than continuous noise.
Unfortunately, this region of the spectrum is outside the range of interferometers since the timescales ($\sim$ 10,000 years) required for a detection are too large, but it may be possible to detect these gravitational waves using, for example, gravitational lensing.

\section{Alternative scenarios}
\label{Nonstandard}

\subsection{Open universes}

There is a substantial body of astronomical evidence which suggests that the
cosmological density parameter is less than critical, $\Omega<1$. Therefore, it
seems sensible to consider the effect of a low-density universe on
the spectrum of gravitational radiation.

The evolution of strings in an open universe will be very much the same as for the
flat case, except that after curvature domination at $t_{\rm curv}\sim \Omega_0t_0$,
where $a(t)\propto t$, the linear scaling regime no longer exists. In fact, it has
been shown\mycite{carlos} that the characteristic length-scale of the network will
increase like $L\sim t [\log(t)]^{1/2}$, rather than $L\sim t$, and hence the long
string density $\rho_{\infty}\propto L^{-2}\propto [t^2\log(t)]^{-1}$, decreases
relative the critical density $\rho_{\rm crit}\propto t^{-2}$, but increases with
respect to the background density $\rho_0\propto t^{-3}$. This is important for
consideration of the gravitational waves created with low frequencies in the matter
era and also for the normalization of the cosmic microwave background\mycite{cald} as
discussed in section \ref{ObservationalBoundsSection}. However, from the point of
view of the red noise spectrum, which is  produced in the radiation dominated era,
there is little difference apart from a shift in the frequency corresponding to
equal matter-radiation.

\begin{figure}
\psfig{figure=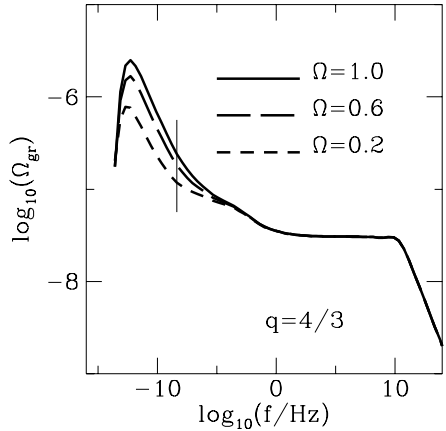,height=7.0cm} 
\caption{
The effect of a low density, $\Omega_0 < 1$ universe on the peaked
portion of the gravitational wave spectrum.  The solid, long- and
short-dashed curves represent spectra for $\Omega_0 = 1, \,0.6,
\,0.2$.  The vertical line shows the location of the frequency bin
probed by pulsar timing measurements. For the loop spectral index
$q=2$, a low density universe dilutes only the lowest frequency waves,
corresponding the radiation emitted by loops still present today.}
\label{figure3a}
\end{figure}

\begin{figure}
\psfig{figure=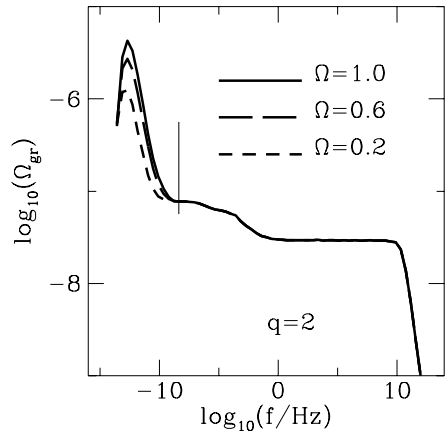,height=7.0cm} 
\caption{
The effect of a low density, $\Omega_0 < 1$ universe on the peaked
portion of the gravitational wave spectrum.  The solid, long- and
short-dashed curves represent spectra for $\Omega_0 = 1, \,0.6,
\,0.2$.  The vertical line shows the location of the frequency bin
probed by pulsar timing measurements. For the loop spectral index
$q=4/3$, a low density universe leads to a dilution of gravitational
waves with wavelengths up to $f \sim 10^{-5}\, {\rm Hz}$. }
\label{figure3b}
\end{figure}

Using methods similar to those used in the previous section, but modified to
accommodate $\Omega_0<1$, we have examined the spectrum of gravitational radiation
produced by the cosmic string network in an open FRW space-time, with $0.1 <
\Omega_0 < 1$. For this range of values of the cosmological density parameter, the
portion of the spectrum produced at a time $t$ is shifted downward by a factor
$\sim \Omega(t)$. Sample spectra for various values of $\Omega_0$ are displayed in
Figures \ref{figure3a}-\ref{figure3b}.  In the case that the spectrum drops off
slower than $1/f$, that is $1<q<2$ and $n_{*} \to \infty$, the spectral density at
frequencies as high as $f \sim 10^{-5}\,{\rm Hz}$ is diluted for $\Omega_0 < 1$. In
the case that the spectrum drops off as $1/f$, that is $q>2$, only at lower
frequencies, $f \lesssim 10^{-10}\,{\rm Hz}$, is the spectral density affected. This
behaviour is exactly as expected: only those frequencies dominated by matter eras
loops are affected.

\subsection{Hybrid defects}

Cosmic strings are generic in a number of symmetry breaking schemes
where the first homotopy group is non-trivial. Simple examples where
this can take place are, when the original symmetry group contains a
$U(1)$ subgroup which is broken to the identity or when the broken
symmetry group has a $Z_2$ subgroup. However, if these simple examples
occur as part of a much more complicated scheme, it is likely that
hybrid systems such as strings connected to either domain walls or
monopoles may form. In the case when the hybrid system does not
annihilate immediately this may lead to a stochastic background in a
similar way to that for strings\mycite{VMa}. We will, therefore,
illustrate the basic physics of this possibility by reference to the
two examples already discussed. The interesting feature of these models
being that they evade the constraints from the cosmic microwave
background and pulsar timing allowing for larger values of $G\mu/c^2$
and hence larger contributions to the stochastic background in the
detectable range of frequencies.

The first and simplest case is that of domain walls connected to strings. This
will occur in the following symmetry breaking scheme,
\begin{equation}
G\rightarrow H \times Z_2 \rightarrow H\,,
\end{equation}
where at the first transition ($T=\eta_{\rm s}$) strings form and at
the second ($T=\eta_{\rm w}$) each string gets attached to a domain
wall. Assuming that the domain walls form during the radiation era,
then before the formation of walls the strings will evolve as in the
standard scenario, creating a nearly flat stochastic background in the
frequency range $f\in [2/\alpha(t_{\rm w}t_{\rm
eq})^{1/2},2/\alpha(t_{*}t_{\rm eq})^{1/2}]$, where $t_{\rm d}={\rm
min}[\mu/\sigma,t_{\rm w}]$ is the time at which the walls dominate
the dynamics of the network, $t_{\rm w}$ is the time of wall formation,
$t_{*}$ is the time when relativistic evolution of the
string network begins and $\mu\sim\eta^2_{\rm s}$, $\sigma\sim\eta_{\rm w}^3$
are the string tension and  wall surface tension respectively. Once the
walls dominate the dynamics of the network, it will break up into what
are effectively string loops spanned by domain walls. These string
loops will collapse into gravitational radiation and other decay
products, such as gauge particles, in about a Hubble time. The exact
nature of this contribution is obviously dependent on the distribution
of the loops produced by the fragmentation process. This process takes place over a relatively short frequency range and may lead to a sharp spike in the spectrum. If a sharp peak in the background spectrum was detected, the the demise of hybrid defects might provide one potential explanation. However, unlike the flat string background, the sensitive dependence on model phenomenology in this case, means that it is not possible at present to predict whether or where such a peak should occur.

The second possibility is of strings connected by monopoles which would have been
created in a symmetry breaking scheme such as,
\begin{equation}
G\rightarrow H \times U(1) \rightarrow H\,,
\end{equation}
where monopoles are formed at the first transition ($T=\eta_{\rm m}$)
and strings connected between monopole/anti-monopole pairs are formed
at the second ($T=\eta_{\rm s}$). In this case there is no period in
which the strings behave as they do in the standard scenario. If the
first transition is after any period of inflation which may have taken
place, then the average separation of the monopoles is less the Hubble
radius and the hybrid system will collapse in one Hubble time by
dissipating energy into friction with the cosmological fluid. However,
if the monopoles are formed during inflation then the system does not
collapse immediately, allowing the formation of a stochastic
background. This scenario has been studied in detail\mycite{VMb} for
the simplest case of a straight string connecting two monopoles. It was
found that the power spectrum emitted by this simple configuration is
divergent with $P_{n}\propto n^{-1}$ due to the perfect symmetry (a
similar spectrum is emitted by a perfectly circular string loop).
Assuming that this divergence is  softened in a less symmetric case,
one may be able to calculate accurately the stochastic background
created. However, it is likely to be dependent on the initial
distribution of monopoles created in the first phase transition.

\section{Observational bounds on the radiation spectrum}
\label{ObservationalBoundsSection}

In this section we determine the observational constraint on the cosmic
string mass-per-unit-length $G\mu/c^2$. To begin, we discuss the recent
analyses of the pulsar timing data, after which we apply the newly
obtained bounds to the cosmic string gravitational wave background.

The observations used to place a limit on the amplitude of a stochastic
gravitational wave background\mycite{KTRref} consist of pulse arrival
times for PSR B1937+21 and PSR B1855+09. Although there has been some
recent  controversy regarding the analysis of the pulsar timing
data\mycite{TDref}, the work by McHugh {\it et al}\mycite{MZVLref} best
assesses the likelihood that the timing residuals are due to
gravitational radiation. We note that all analyses to date have assumed
a flat, red noise  spectrum for the gravitational wave spectral
density. Such an assumption  is only justified for a restricted range
of frequencies in the case of a background due to cosmic strings, as we
have demonstrated in the  preceding section. Hence, a statistical
analysis which uses a realistic  model of the cosmic string spectrum
may obtain a different limit on the  amplitude of the spectral
density.

\begin{figure}
\psfig{figure=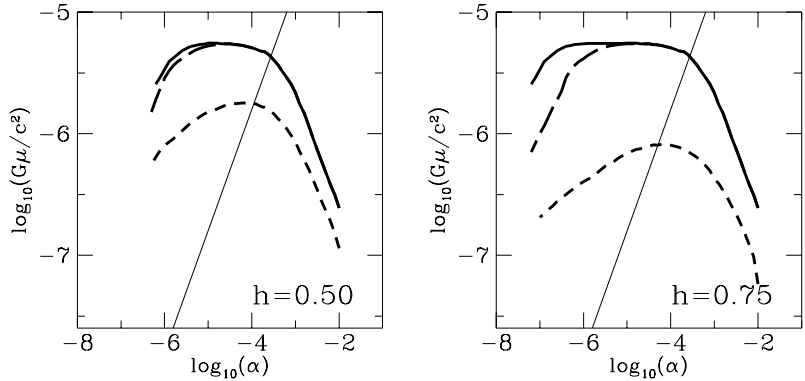,height=6.0cm} 
\caption{
Curves of constant $\Omega_{\rm gr}$ in $(\alpha, G\mu/c^2)$ parameter space are
shown. For a given value of $\alpha$, these figures give the observational bound on
$G\mu/c^2$ in the case $h=0.5, \, 0.75$. In each figure, the constraining curves for
$q=10,\, 2,\, 4/3$ are given by the solid, long-, and short-dashed curves. The light
dashed lines show $\alpha = \Gamma G\mu/c^2$. The most conservative constraint is
$G\mu/c^2 < 5.4 \times 10^{-6}$.  }
\label{figure4}
\end{figure}

We now present values of the parameter $G\mu/c^2$ for values of
$\alpha$ which satisfy the pulsar timing constraint on the
gravitational radiation spectrum.  Contours of constant $\Omega_{\rm
gr}$ in the logarithmic frequency bin $f = (8\, {\rm yrs})^{-1}$, given
by (\ref{GRBLimits}), in $(\alpha,G\mu/c^2)$ parameter space, are shown
in Figure \ref{figure4}. We have used cosmological parameters $\Omega_0
= 1$ and $h \in [0.5,\,0.75]$ with the cosmic string loop radiation
efficiency $ \Gamma = 60$.   We find
\begin{equation}
G\mu/c^2 < \cases{
2.0 (\pm 0.4) \times 10^{-6} \,\, (2 h)^{-8/3}	& $q = 4/3$ \cr
5.4 (\pm 1.1) \times 10^{-6}	& $q \ge 2$ or $n_{*} \lesssim 10^{2}$.}
\label{pulsarbound}
\end{equation}
These constraints correspond to the maximum value of $G\mu/c^2$ along
the contour of constant  $\Omega_{\rm gr}$.  In the case $q=4/3$, this
maximum occurs near $\alpha =  \Gamma  G \mu/c^2$, the expected size of
newly formed loops based on considerations of the gravitational
back-reaction, while for the $q \ge 2$ or $n_{*} \lesssim 10^{2}$ case,
the maximum occurs at a slightly smaller value of $\alpha$.  For both
larger and smaller values of $\alpha$ the bounds become more
stringent\mycite{RadiationCalcref}.  The unusual dependence on $h$ is
due to the contribution from high mode number waves emitted in the
matter era, for which the amplitude depends on both the slope of the
spectrum and the time of radiation-matter equality.  The quoted errors
are due to uncertainties in the cosmic string model parameters measured
by the numerical simulations\mycite{Simulationref}.  For the case of an
open universe, there is no change in the $q \ge 2$ or $n_{*} \lesssim
10^{2}$ bound. However, the $q=4/3$ bound is weakened by a factor $\sim
1/\Omega_0$.

We now comment on the validity of the model which we have used to
generate the gravitational radiation spectra.  We have shown that the
observational bounds on the total spectrum are sensitive to the value
of the loop spectral index $q$, unless there is a back-reaction cut-off
$n_{*} \lesssim 10^2$.  Furthermore, we have noted that there is
uncertainty in the characteristic value of the loop spectral index,
$q$.  Hence, we feel that it is more reasonable to take the
conservative bound of (\ref{pulsarbound}) at the present.  Next,
consider the extended one-scale
model\mycite{Reviewref,RadiationCalcref}, for the evolution of the
string network. This model assumes that the long string energy density
scales relative to the background energy density, with the dominant
energy loss mechanism due to the formation of loops of a characteristic
scale.  A more sophisticated model, by Austin {\it et
al} \mycite{ACKModelref}, attempts to include the effect of the
gravitational back-reaction on the long-term evolution of the string
network; results suggest that an effect of the back-reaction may be to
lower the scaling density in long strings at late times, beyond the
reach of numerical simulations. Hence, there is some uncertainty as to
how accurately the extended one-scale model describes the evolution of
the string network.  However, we do not believe that these
considerations could result in a decrease in the amplitude of the
gravitational wave background by more than $\sim 50\%$.  Thus, we quote
$G\mu /c^2 < 5.4 (\pm 1.1) \times 10^{-6}$ as a conservative bound on
the cosmic string mass-per-unit-length.

The bounds already computed\mycite{RadiationCalcref} due to the
constraint on the total energy density in gravitational waves at the
time of nucleosynthesis remain valid. For a limit on the effective
number of neutrino species $N_\nu < 3.1,\, 3.3,\, 3.6$, the bound on
the cosmic string mass-per-unit-length is $G\mu/c^2 < 2,\, 6,\, 10
\times 10^{-6}$ respectively, evaluated at $\alpha = \Gamma G\mu/c^2$.
The big-bang nucleosynthesis limit on the number of effective neutrino
species is a conservative $N_\nu < 4$, owing to uncertainties in the
systematic errors in the observations of light element
abundances\mycite{BBNLimitsref}.  Hence, until the observations are
refined, the nucleosynthesis bound is weaker than the pulsar timing
bound. Furthermore, the translation of the limit on $N_\nu$ into the
bound on $G\mu/c^2$ is sensitive to the thermal history of the
cosmological fluid\mycite{BennettPapersref}. The bound on the string
mass-per-unit-length may be considerably weakened if the cosmological
fluid possessed many more relativistic degrees of freedom in the early
universe beyond those given by a minimal GUT model.

Comparing detailed computations of the large angular scale cosmic
microwave background temperature anisotropies induced by cosmic
strings\mycite{CMBref} with observations, the cosmic string
mass-per-unit-length has been normalized to
\begin{equation} 
G\mu/c^2 = 1.05^{\,+0.35}_{\,-0.20}\, \times 10^{-6}.
\label{NormalizationEquation} 
\end{equation} 
In an open universe, this normalization is expected to increase
slightly\mycite{cald} to $G\mu/c^2 \sim 1.6 \times 10^{-6}$ for
$\Omega_o = 0.1$.  Therefore, given the uncertainties in the extended
one-scale model, we find the gravitational radiation spectrum to be
compatible with observations.

We have already noted that the hybrid systems evade the constraint from
pulsar timing unless they survive until around the time of
nucleosynthesis $(t_{\rm nucl}\sim 1{\rm sec})$, much later than thought
possible according to the standard model of particle physics.
Therefore, the only constraint on the stochastic background comes from
Big Bang nucleosynthesis. The interesting thing to note is that since
the hybrid network will annihilate at some time $t_{\rm a}$, well
before nucleosynthesis, then assuming that the spectrum is flat, the
amplitude of the stochastic background allowed is larger by a factor of
\begin{equation}
\log\left({t_{\rm nucl}\over t_{*}}\right)\bigg/\log\left({t_{\rm a}\over 
t_{*}}\right)\,.
\end{equation}
For the first transition at the GUT scale, that is $t_{*}\sim
10^{-32}$, and the second around $10^{10}{\rm GeV}$ with almost
immediate annihilation, then this results in the constraint being
weakened by approximately a factor of 10, corresponding to $G\mu\sim
10^{-5}$. This makes a detection in the first generation of LIGO
detectors marginally possible\mycite{VMa}, although the parameter values
would have to be rather extreme.

\section{Detection of the Radiation Spectrum} \label{DetectionSection}

We would like to determine whether the stochastic gravitational wave
spectrum emitted by cosmic strings may be observed by current and
planned detectors.  Because all ground-based detectors operate at
frequencies $f \gtrsim 10^{-3}\, {\rm Hz}$, we need only consider the
`red noise' portion of the gravitational wave spectrum
(\ref{RedNoiseSpectrumToday}). Noting that the spectral density,
$\Omega_{\rm gr}(f)$, has a minimum value when $\alpha \to 0$ the predicted
spectrum is bounded from below by
\begin{eqnarray}
\Omega_{\rm gr}(f) &\ge &  {24 \pi \over 9} A f_{\rm r}
{(1 - \langle v^2 \rangle/c^2 ) \over (1 + z_{\rm eq})} (G\mu/c^2)
\Bigl({g(T_0) / g(T_{\rm GUT})}\Bigr)^{1/3} \cr\cr
&\ge & 1.4 \times 10^{-9} \qquad
{\rm for}\quad 10^{-8}\, {\rm Hz} \lesssim f \lesssim 10^{10}\, {\rm Hz}.
\label{MinimumSpectrum}
\end{eqnarray}
Here we have used the normalization in (\ref{NormalizationEquation})
for $G\mu/c^2$, Hubble parameter $h = 0.75$, and assumed a minimal GUT
thermal history.  Hence, this lower bound is valid up to frequencies $f
\sim 10^{-3} \alpha^{-1}\,{\rm Hz} \sim 10 \, {\rm Hz}$ based on our
knowledge of the number of relativistic degrees of freedom, $g$, of the
primordial fluid up to temperatures $T \sim 10^3 \, {\rm GeV}$.  Notice
that a measurement of the spectral density due to cosmic strings at
higher frequencies would sample $g$ at higher temperatures.

We may in turn place a lower bound on the amplitude of the
dimensionless strain predicted for the gravitational wave emitted by
cosmic strings:
\begin{eqnarray}
h_{\rm c} &=& 1.3 \times 10^{-18} h \sqrt{\Omega_{\rm gr}(f)}
\Bigl({f \over 1 \, {\rm Hz}}\Bigr)^{-1} \cr\cr
&\ge &3.6 \times 10^{-23}
\Bigl({f \over 1 \, {\rm Hz}}\Bigr)^{-1}
\qquad
{\rm for}\quad 10^{-6}\, {\rm Hz} \lesssim f \lesssim 10^{8}\, {\rm Hz}.
\label{MinimumStringStrain}
\end{eqnarray}
The expressions (\ref{MinimumSpectrum}-\ref{MinimumStringStrain}) are
useful for comparison with the planned sensitivities of the forthcoming
generation of gravitational wave detectors\mycite{AmaldiConfref}.

The most promising opportunity to probe for a stochastic gravitational
wave background due to cosmic strings is through a cross-correlation of
the observations of the advanced LIGO, VIRGO and LISA interferometers.
It is estimated that the advanced LIGO detectors will have the sensitivity
\begin{equation}
h_{\rm 3/yr} = 5.2 \times 10^{-25} \Bigl({f \over 1 \, {\rm k Hz}}\Bigr)^{1/2}
\end{equation}
for stochastic waves (equation 125c of Thorne\mycite{LIGOref}), sufficient to
measure the minimum predicted strain (\ref{MinimumStringStrain}) near
$f \sim 100 \,{\rm Hz}$ in a $1/3$-year integration time.  More recent
calculations\mycite{LIGOPosref} confirm that the orientation of the
advanced LIGO interferometers will be sufficient in order to detect the
cosmic string gravitational wave background. For the LISA
project\mycite{LISAref}, comparison of the projected strain sensitivity
$h_c \sim 10^{-20}$ at the frequency  $f \sim 10^{-3}\, {\rm Hz}$ with
(\ref{MinimumStringStrain}) indicates that the space-based
interferometer will be capable of detecting a gravitational radiation
background produced by a network of cosmic strings.

Other ground-based interferometric gravitational wave detectors are in
development or under construction. The GEO600 and TAMA300 detectors,
operating near frequencies $f \sim 10^3 \, {\rm Hz}$, may also be
capable of measuring a cosmic string generated background.

A network of resonant mass antennae, such as bar and TIGA detectors may
probe for a stochastic background.  Successful detection by these
antennae will require improved sensitivity and longer integration
time.  However, cross-correlation between a narrow-band bar and a
wide-band interferometric detector may improve the opportunities.
Estimates of the sensitivity of such a system, assuming optimum
detector alignment\mycite{Correlationref}, indicate that
\begin{equation}
\sqrt{h_{\rm int} h_{\rm b}} \gtrsim
2.6 \times 10^{-19} \sqrt{\Omega_{\rm gr}(f)}
\Bigl({f \over 1 \, {\rm k Hz}}\Bigr)^{-3/2}
\Bigl({t_{\rm obs} \over 10^7 \, {\rm s}}\Bigr)^{1/2}
\end{equation}
is necessary to detect a background $\Omega_{\rm gr}$.  Hence, for a
$1/3$-year observation time, the bar and interferometer strain
sensitivities at $1 \, {\rm kHz}$ must be better than $\sim 10^{-23}$
in order to detect the cosmic string background.

We stress that the amplitude of the cosmic string gravitational wave
background for frequencies  $f \gtrsim 10\,{\rm Hz}$ is sensitive to
the number of degrees of freedom of the cosmological fluid at
temperatures $T \gtrsim 10^3 \,{\rm GeV}$. The amplitude of the cosmic
string background at LISA-frequencies, near $f \sim 10^{-3}\, {\rm
Hz}$, is firm, since the cosmological fluid near the temperature $T
\sim 10 \,{\rm MeV}$ is well understood.  However, at the higher
frequencies probed by ground-based detectors, our uncertainty in the
number of degrees of freedom of the cosmological fluid, as determined
by the correct model of particle physics at that energy scale, may
reduce the predicted amplitude (\ref{MinimumSpectrum}) of gravitational
radiation.

\begin{figure}
\centerline{\psfig{figure=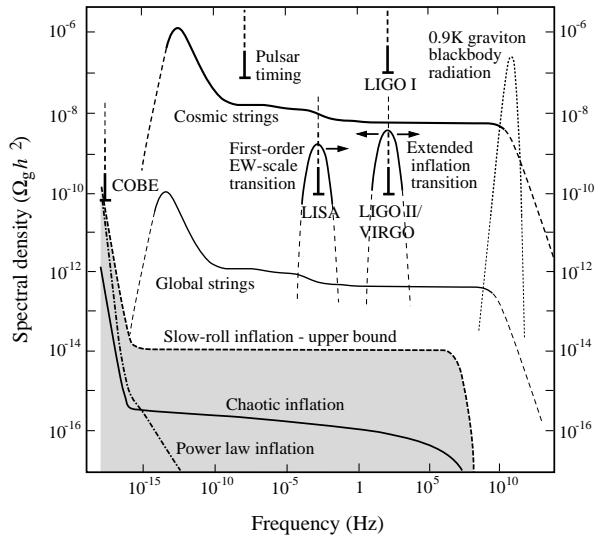,width=14.0cm}}
\caption{Summary of the potential cosmological sources of a stochastic gravitational
radiation background, including inflationary models, first order phase transitions
and cosmic strings, as well as a primordial 0.9K blackbody graviton spectrum (the
analogue of the blackbody photon radiation). Also plotted are the relevant
constraints from the COBE measurements, pulsar timings, and the sensitivities of the
proposed interferometers. Notice that local cosmic strings and strongly first-order
phase transitions may produce detectable backgrounds in contrast to standard
slow-roll inflation models.}
\label{figure5}
\end{figure}


\section{Conclusion}
\label{ConclusionSection}

We have presented improved calculations of the spectrum of relic
gravitational waves emitted by cosmic strings.  We demonstrated that
the effect of a gravitational back-reaction on the radiation spectrum
of cosmic string loops, for which there is an effective mode cut-off
$n_{*} \lesssim 10^2$, may serve to weaken the pulsar timing bound on
the cosmic string mass-per-unit-length.  Arguing for a model of
radiation by loops, for which either the spectral index is $q \ge 2$ or
there is an emission mode cut-off $n_{*} \lesssim 10^2$, we obtain the
conservative bound $G \mu/c^2 < 5.4 (\pm 1.1) \times 10^{-6}$ due to
observations of pulsar timing residuals. We believe this bound to be
robust, in that the spectrum depends weakly on the precise value of the
mode cut-off, up to $n_{*} \sim 10^2$. 
We have noted the interesting
result that the flat, red noise portion of the gravitational wave
spectrum is sensitive to the thermal history of the cosmological fluid,
revealing features of the particle physics content at early times. We
have also discussed the possibility of a low-$\Omega$ universe and
hybrid systems of defects, such as strings connected by domain walls or
monopoles.  Finally, we have pointed out that the generation of
advanced LIGO, VIRGO and LISA interferometers should be capable of
detecting the predicted stochastic gravitational wave background due to
cosmic strings.

We may place the cosmic string scenario in context with other candidate
sources for a stochastic background of cosmological gravitational
waves\mycite{BS} in Fig. \ref{figure5}.  These include gravitational
waves created during inflation\mycite{tensor} and through bubble
collisions following a first-order phase transition\mycite{first}.
Viable theoretical scenarios within detector sensitivity include
strongly first-order phase transitions, possibly at the end of
inflation, and networks of cosmic strings. At this stage, other
primordial backgrounds from slow-roll inflation, global topological
defects and the standard electroweak phase transition appear to be out
of range. The discovery of any of the possible cosmological sources
will have enormous implications for our understanding of the very early
universe and for fundamental physics at the highest energies.

\section*{Acknowledgments}
The work of EPSS is supported by PPARC grant GR/K29272. The work
of RAB is supported by PPARC postdoctoral fellowship grant GR/K94799.
The work of RRC is supported by the DOE at Penn (DOE-EY-76-C-02-3071).



\end{document}